\documentclass[10pt,twocolumn,pra,showpacs,amsmath,amssymb]{revtex4}

\usepackage{mathrsfs}
\usepackage[english]{babel}  
\usepackage{amsfonts}
\usepackage{tensor}

\def\bra#1{\mathinner{\langle{#1}|}}
\def\ket#1{\mathinner{|{#1}\rangle}}
\def\braket#1{\mathinner{\langle{#1}\rangle}}
\def\Bra#1{\left\langle#1\right|}
\def\Ket#1{\left|#1\right \rangle}
{\catcode`\|=\active 
  \gdef\Braket#1{\begingroup
\mathcode`\|32768\let|\BraVert\left<{#1}\right>\endgroup}}
\def\BraVert{\egroup\,\mid\,\bgroup}
\def\Brak#1#2#3{\bra{#1}#2\ket{#3}}
\usepackage{cancel}
\usepackage{color}
\definecolor{Blue}{rgb}{0,0,1}
\definecolor{Red}{rgb}{1,0,0}
\definecolor{Green}{rgb}{0,1,0}
\definecolor{Purp}{rgb}{.2,0,.2}
\definecolor{white}{rgb}{1,1,1}

\begin{document}

\title{How state preparation can affect a quantum experiment: Quantum process tomography for open systems}

\author{Aik-meng Kuah}
 \affiliation{Center for Complex Quantum Systems, The University of Texas at Austin, Austin Texas 78712}
\author{Kavan Modi}
 \email{modik@physics.utexas.edu}
 \affiliation{Center for Complex Quantum Systems, The University of Texas at Austin, Austin Texas 78712}
\author{C\'{e}sar A. Rodr\'{i}guez-Rosario} 
 \affiliation{Center for Complex Quantum Systems, The University of Texas at Austin, Austin Texas 78712}
\author{E.C.G. Sudarshan}
 \affiliation{Center for Complex Quantum Systems, The University of Texas at Austin, Austin Texas 78712}

\date{\today}

\begin{abstract}
We study the effects of preparation of input states in a quantum tomography experiment.  We show that maps arising from a quantum process tomography experiment (called process maps) differ from the well know dynamical maps.  The difference between the two is due to the preparation procedure that is necessary for any quantum experiment. We study two preparation procedures, stochastic preparation and preparation by measurements. The stochastic preparation procedure yields process maps that are linear, while the preparations using von Neumann measurements lead to non-linear processes, and can only be consistently described by a bi-linear process map.  A new process tomography recipe is derived for preparation by measurement for qubits.  The difference between the two methods is analyzed in terms of a quantum process tomography experiment.  A verification protocol is proposed to differentiate between linear processes and bi-linear processes.  We also emphasize the preparation procedure will have a non-trivial effect for any quantum experiment in which the system of interest interacts with its environment.
\end{abstract}


\maketitle

%%%%%%%%%%%%%%%%%%%%%%%%%%%%%%%%%%%%%%%%
%%%%%%%%%%%%%%%%%%%%%%%%%%%%%%%%%%%%%%%%
\section{Introduction}
%%%%%%%%%%%%%%%%%%%%%%%%%%%%%%%%%%%%%%%%
%%%%%%%%%%%%%%%%%%%%%%%%%%%%%%%%%%%%%%%%

Quantum process tomography \cite{JModOpt.44.2455, PhysRevLett.78.390} is the experimental tool that determines the open evolution of a system that interacts with the surrounding environment. It is an important tool in the fields of quantum computation and quantum information that allows an experimenter to determine the action of a quantum process on the system of interest.

The standard tomography procedure and some variations of it (namely entanglement assisted tomography \cite{PhysRevLett.86.4195, PhysRevA.67.062307}, ancilla assisted tomography \cite{PhysRevLett.90.193601,PhysRevLett.91.047902}, and direct characterization of quantum dynamics \cite{mohseni:170501,mohseni:062331}) have been verified experimentally \cite{Nielsen:1998py,PhysRevA.64.012314,PhysRevLett.91.120402,Wein:121.13,orien:080502,Howard06,myrskog:013615}.  

In many of these experiments, the maps that characterize the quantum process have been plagued with negative eigenvalues and sometimes non-linear behavior.  It was pointed out previously that the negative eigenvalues may be due to the initial correlations between the system and the environment \cite{JordanShajiSudarshan04, CesarEtal07, StelmachovicBuzek01, CarteretTernoZyczkowski05, Ziman06}.

Quantum process tomography is thought of as a procedure that allows us to experimentally determine the dynamical map describing the quantum process.  However, all experiments require a method to prepare the initial states of the system at the beginning of the experiment.  This act of preparation has been neglected from the theory of quantum process tomography.  

We will show in this paper how an open system is prepared into different initial states can fundamentally change the outcome.  A consequence of our study will force us to incorporate state preparation into the map that describes the process.  We will have to distinguish between maps determined from a quantum process tomography experiment, which we will call \emph{process maps}, from the well know \emph{dynamical maps} \cite{SudarshanMatthewsRau61,SudarshanJordan61}.  The key difference between dynamical maps and process maps is that process maps include the initial step of state preparation, while dynamical maps are not restricted in that sense (see Appendix \ref{pvd}).

We will study two methods for preparing states for quantum experiments, the stochastic preparation method and preparations using von Neumann measurements.  In the former case, the process is given by a completely positive linear map (also see \cite{Ziman06} for an independent, but related discussion).  However for the measurement method, we will show that the outcome of the experiment cannot be consistently described by a linear map.  We propose a bi-linear process map to describe such a quantum process, and we will show that this bi-linear map can be experimentally determined by developing a procedure for bi-linear quantum process tomography.

Based on our results existing quantum process tomography experiments should be reanalyzed.  Any experiment which obtained a process map that had negative eigenvalues or behaved in non-linear fashion may suffer from poor preparation of input states, and should be analyzed for bi-linearities.

%%%%%%%%%%%%%%%%%%%%%%%%%%%%%%%%%%%%%%%%
%%%%%%%%%%%%%%%%%%%%%%%%%%%%%%%%%%%%%%%%
\section{Linear Quantum Process Tomography}\label{LQPT}
%%%%%%%%%%%%%%%%%%%%%%%%%%%%%%%%%%%%%%%%
%%%%%%%%%%%%%%%%%%%%%%%%%%%%%%%%%%%%%%%%

The objective of quantum process tomography is to determine how a quantum process acts on different states of the system.  In very basic terms, a quantum process takes different quantum input states to different output states:
\begin{equation*}
\text{Input states} \rightarrow \text{PROCESS} \rightarrow \text{Output states}.
\end{equation*}
The complete behavior of the process is known if the output state for any given input state can be predicted.

The tomography aspect of quantum process tomography, is to use a finite number of carefully selected input states instead of all possible states, to determine the process.  If the quantum process is described by a linear map (see \cite{Nielsen00a} for detailed discussion), then the necessary input states should linearly span the state space of the system.  For a finite dimensional state space, this requires a finite number of input states.  Once the evolution of these input states is known, then by linearity the evolution of any input state is known.  

For example for a qubit only the following four projections as input states are necessary
\begin{eqnarray}\label{prj}
P^{(1,-)} = \frac{1}{2}(\openone - \sigma_1),\;
P^{(1,+)} = \frac{1}{2}(\openone + \sigma_1 ),\nonumber\\
P^{(2,+)} = \frac{1}{2}(\openone + \sigma_2),\;\;
\mbox{and}\;\;P^{(3,+)} = \frac{1}{2}(\openone + \sigma_3)
\end{eqnarray}
to linearly span the whole state space, i.e. any state of a qubit can be written as a unique linear combination of these four projections.  Above, $\openone$ is the $2 \times 2$ identity matrix and $\sigma_j$ are the Pauli spin matrices.  

Using the set linearly independent input states $P^{(j)}$, and measuring the corresponding output states $Q^{(j)}$, the evolution of an arbitrary input state can be determined.  Let the linear map describing the process be given by $\Lambda$, and the arbitrary input state be expressed (uniquely) as a linear combination $\sum_j p_j P^{(j)}$.  Then the action of the map in terms of the matrix elements is as follows:
\begin{eqnarray*}
\sum_{r's'}\Lambda_{rr',ss'}\left(\sum_j p_j P_{r's'}^{(j)}\right) 
&=& \sum_j p_j Q_{rs}^{(j)}.
\end{eqnarray*}
The map itself can be expressed as
\begin{equation}\label{qptlin}
\Lambda_{rr',ss'} = \sum_n Q_{rs}^{(n)} {\tensor*{\tilde{P}}{*^{(n)}_{r's'}}}^*,
\end{equation}
where $\tilde{P}^{(n)}$ are the duals of the input states satisfying the scalar product 
\begin{eqnarray*}
{\tensor{\tilde{P}}{^{(m)}}}^{\dag}P^{(n)}=\sum_{rs}{\tensor*{\tilde{P}}{*^{(m)}_{rs}}}^*{P^{(n)}_{rs}} = \delta_{mn}.
\end{eqnarray*}
The duals for the projections in Eq. (\ref{prj}) are
\begin{eqnarray}\label{dual}
\tilde{P}^{(1,-)} &=& \frac{1}{2}(\openone - \sigma_1- \sigma_2- \sigma_3),\nonumber\\
\tilde{P}^{(1,+)} &=& \frac{1}{2}(\openone + \sigma_1-\sigma_2-\sigma_3 ), \\
\tilde{P}^{(2,+)} &=&\sigma_2,\;\;\;\mbox{and}\;\;\;
\tilde{P}^{(3,+)} =\sigma_3.\nonumber
\end{eqnarray}

%%%%%%%%%%%%%%%%%%%%%%%%%%%%%%%%%%%%%%%%
%%%%%%%%%%%%%%%%%%%%%%%%%%%%%%%%%%%%%%%%
\section{Preparation of Input States}\label{prep}
%%%%%%%%%%%%%%%%%%%%%%%%%%%%%%%%%%%%%%%%
%%%%%%%%%%%%%%%%%%%%%%%%%%%%%%%%%%%%%%%%

It is important to note the assumption made at the beginning of the last section, that is, the process is given by a linear map.  This assumption is made without any consideration to how the input states are prepared.  We will show that for the same input state, evolving through the same Hamiltonian, the output state would be different depending on whether the input state is prepared stochastically or by a measurement. 

The basic steps in a quantum process tomography experiment are broken down below:
\begin{itemize}
\item[$i.$]{Just before the experiment begins, the system and environment is in an unknown state, which we will write as $\gamma_0$.  The system and environment could be entangled or correlated.  For our discussions we will label the system as $\mathbb{A}$ and the environment as $\mathbb{B}$.}  
\item[$ii.$] {The system is prepared into a known input state.  Let $\mathscr{P}^{(n)}$ be the map that prepares the system into the $n^{th}$ input state.  The system and environment state after preparation is therefore given by $(\mathscr{P}^{(n)} \otimes \mathcal{I})(\gamma_0)$, where $\mathcal{I}$ is the identity map acting on the space of the environment.}
\item [$iii.$]{The system is then sent through an unknown quantum process.  We consider the evolution to be a unitary transformation $U$ in the space of the system \emph{and} environment:
\begin{eqnarray*}
U (\mathscr{P}^{(n)}\otimes\mathcal{I})(\gamma_0) U^\dag.
\end{eqnarray*}}
\item [$iv.$]{The trace with respect to the environment is taken to obtain the output state of the system
 \begin{eqnarray}\label{output}
 Q^{(n)}=\mbox{Tr}_{\mathbb B}\left[U (\mathscr{P}^{(n)}\otimes\mathcal{I})(\gamma_0) U^\dag \right],
 \end{eqnarray}
\noindent where $Q^{(n)}$ is the output state corresponding to the input state prepared by $\mathscr{P}^{(n)}$.}
\item [$v.$]{Finally using the input and the output states, a map describing the process is constructed.}
\end{itemize}

It is important to keep in mind that these basic steps (excluding the last step) also describe most quantum experiments, not just specifically quantum process tomography experiments.  Therefore the following results may be applicable to many quantum experiments, not just to quantum process tomography experiments.   

%%%%%%%%%%%%%%%%%%%%%%%%%%%%%%%%%%%%%%%%
%%%%%%%%%%%%%%%%%%%%%%%%%%%%%%%%%%%%%%%%
\section{Stochastic Preparation}\label{stoprep}
%%%%%%%%%%%%%%%%%%%%%%%%%%%%%%%%%%%%%%%%
%%%%%%%%%%%%%%%%%%%%%%%%%%%%%%%%%%%%%%%%

There are two methods in quantum theory to prepare an unknown state into a known state;  stochastic preparations and preparations by measurements.  In this section, we will discuss the method of using stochastic maps for preparation of states.

Let us consider preparation by a stochastic pin map which maps all states to a fixed single state \cite{gorini76}.  Using a pin map for preparation is a common initial step in various experiments.  For example, in quantum dot experiments the system is cooled very close to absolute zero temperature.  This ensures the probability of the system being in the ground energy state is nearly one.  This is effectively a pin map to the ground energy state.

The experiment procedure will begin with a pin map $\Theta$, which takes any density matrix to a fixed pure state $\ket{\Phi}$.  Then the state of the system and environment after the pin map is:
\begin{equation}\label{pinmap}
\left(\Theta \otimes \mathcal{I}\right) \left(\gamma_0\right) = \Ket{\Phi}\Bra{\Phi} \otimes \tau (\Theta),
\end{equation}
where $\tau(\Theta) = \mbox{Tr}_\mathbb{A}[(\Theta \otimes \mathcal{I}) \left(\gamma_0\right)]$. The pin map fixes the system into a single pure state, which in turn means that the state of the environment is fixed into a single state as well.  The purpose of the pin map is to decouple the system from the environment, to eliminate any correlation between the system state and the environment state.  Note that the state of the environment does depend on the choice of the pin map $\Theta$.  

Once the pin map is applied, the system has to be prepared into the various different input states for the tomography experiment.  This can be expressed in the most general way with stochastic maps:
\begin{eqnarray}\label{stocmap}
\Omega^{(n)} \left( \Ket{\Phi}\Bra{\Phi} \right) = P^{(n)},
\end{eqnarray}
where $P^{(n)}$ are the desired input states.

The preparation procedure can be summarized as $\mathscr{P}^{(n)} =\Omega^{(n)} \circ\Theta $.  The overall experiment can be written in a single equation by combining Eqs. (\ref{output}$-$\ref{stocmap}):
\begin{eqnarray}\label{stocproceq}
Q^{(n)} 
& =& \mbox{Tr}_\mathbb{B} \bigg{[} U \left(\left[\Omega^{(n)} \circ\Theta\right] \otimes \mathcal{I} \right) \left(\gamma_0\right) U^\dagger\bigg{]}\nonumber \\
& =& \mbox{Tr}_\mathbb{B}\left[U P^{(n)} \otimes \tau(\Theta) U^\dagger\right].
\end{eqnarray}

We will call this equation `the process equation'. Notice that this process equation is linear on $P^{(n)}$. 
Once the input states are prepared, the procedure for quantum process tomography is the same as given in section \ref{LQPT}.  This is a generalization of linear quantum process tomography.

It should be emphasized that the initial pin map $\Theta$ is critical; because for the process to be linear the state of the environment must be independent of the input state.  It may be tempting to simply use a set of pin maps, $\Theta^{(n)}$, to prepare the various input states $P^{(n)}$.
However, the process equation in this case yields:
\begin{eqnarray}
Q^{(n)} 
& =& \mbox{Tr}_\mathbb{B} \bigg{[} U \left. \Theta ^{(n)} \otimes \mathcal{I} \right. \left(\gamma_0\right) U^\dagger\bigg{]}\nonumber \\
& =& \mbox{Tr}_\mathbb{B}\left[U P^{(n)} \otimes \tau(\Theta ^{(n)}) U^\dagger\right].
\end{eqnarray}
This is no longer a linear equation on $P^{(n)}$, since the state of the environment, $\tau(\Theta ^{(n)})$, is effectively dependent on $P^{(n)}$.

%%%%%%%%%%%%%%%%%%%%%%%%%%%%%%%%%%%%%%%%
\subsection{Example of Stochastic Preparation}
%%%%%%%%%%%%%%%%%%%%%%%%%%%%%%%%%%%%%%%%
It may instructive to look at a simple example involving two qubits at this point.  We will treat one qubit as the system of interest and the other as the unknowable environment.

Consider the preparation by a pin map $\Theta$
\begin{eqnarray}\label{inp1}
\Theta\otimes\mathcal{I}\left(\gamma_0\right)=\ket{\Phi}\bra{\Phi}\otimes\frac{1}{2}\openone.
\end{eqnarray}
that yields a pure state $\ket{\Phi}$ for the system qubit and a completely mixed state for the environment qubit. 

The next step is to create different input states using different maps $\Omega^{(n)}$.  In this case, the fixed state $\ket{\Phi}\bra{\Phi}$ can simply be locally rotated to get the desired input state $P^{(n)}$ given in Eq. (\ref{prj})
\begin{eqnarray}\label{inp2}
\Omega^{(n)}\left(\ket{\Phi}\bra{\Phi}\right)\otimes\frac{1}{2}\openone&=& V^{(n)}\ket{\Phi}\bra{\Phi}{V^{(n)}}^{\dag}\otimes\frac{1}{2}\openone\nonumber\\
&=&P^{(n)}\otimes\frac{1}{2}\openone,
\end{eqnarray}
where $n=\{(1,-),(1,+),(2,+),(3,+)\}$ and $V^{(n)}$ are the unitary operators acting on the space of the system.  Now each input state is sent through the quantum process. The output states can be calculated in a straight forward manner using the process equation Eq. (\ref{stocproceq}).
For this example we chose the unitary transformations $U$ to be 
\begin{eqnarray}\label{unitary}
U=e^{-i Ht}=e^{-i\sum_j\sigma_j\otimes\sigma_j\; t}. 
\end{eqnarray}
The output states are
\begin{eqnarray*}
Q^{(1,-)}=\frac{1}{2}(\openone-C^2\sigma_1),&&
Q^{(1,+)}=\frac{1}{2}(\openone+C^2\sigma_1),\\
Q^{(2,+)}=\frac{1}{2}(\openone+C^2\sigma_2),
 &\mbox{and}&Q^{(3,+)}=\frac{1}{2}(\openone+C^2\sigma_3).
\end{eqnarray*}

The linear process map is constructed using Eq. (\ref{qptlin}), the duals in Eq. (\ref{dual}), and the output states:
\begin{eqnarray*}
\Lambda_s=\frac{1}{2}
\left(\begin{array}{cccc}
1+C^2 &0&0&2C^2\vspace{.2cm}\\
0& 1-C^2& & 0 \vspace{.2 cm}\\
0 & 0& 1-C^2 &0 \vspace{.2 cm}\\
2C^2& 0&0& 1+C^2 \vspace{.2 cm}
\end{array}\right),
\end{eqnarray*}
where $C=\cos(2t)$.

We can verify that the process is correctly represented by this linear map by calculating $\Lambda_s (\rho)$ and comparing it to a direct calculation $\mbox{Tr}_\mathbb{B}[U \rho \otimes \frac{1}{2} \openone U^\dagger]$, for any $\rho$.

%%%%%%%%%%%%%%%%%%%%%%%%%%%%%%%%%%%%%%%%
%%%%%%%%%%%%%%%%%%%%%%%%%%%%%%%%%%%%%%%%
\section{Preparation by Measurements}\label{mespreps}
%%%%%%%%%%%%%%%%%%%%%%%%%%%%%%%%%%%%%%%%
%%%%%%%%%%%%%%%%%%%%%%%%%%%%%%%%%%%%%%%%

Making a quantum measurement is another effective method of preparing input states for an experiment.  For our discussion, we will use the von Neumann model for the measurement process.  With the von Neumann model, a measurement is given by a set of orthonormal projections. If a particular outcome is observed from the measurement, the state of the system collapses to that corresponding projection.  Therefore, the input states can be prepared for an experiment by suitably fixing our measurement basis. With the knowledge of the basis and the outcomes, the exact input state is known.

For experiments dealing with polarization states of photons this is a very practical method for preparation.  Sending a randomly polarized beam of photons through a polarizer will yield in a perfectly polarized beam of light.  This polarized beam can e used as one of the input state for a quantum tomography experiment.

Suppose the $n^{th}$ input state is prepared, given by the projection $P^{(n)}$, by measurement.  For the open system, the state becomes
\begin{equation}\label{mesprep}
\mathscr{P}^{(n)}\otimes\mathcal{I}(\gamma_0)= \frac{1}{\Gamma_n} \left(P^{(n)}\otimes\openone\right) \gamma_0  \left(P^{(n)}\otimes\openone\right),
\end{equation}
where $\Gamma^{(n)} = \mbox{Tr}[ \left(P^{(n)}\otimes\openone\right) \gamma_0]$ is the normalization factor (it is the probability of obtaining that particular input state from a von Neumann measurement).  For simplicity, we will from this point on, simply write $P^{(n)}$ instead of $P^{(n)}\otimes\openone$.

Combining Eq. (\ref{output}) and Eq. (\ref{mesprep}) leads to the process equation that will relate the output states to the input states. For the $n^{th}$ input state the process equation is
\begin{equation}\label{BasicProcessEquation}
Q^{(n)} = \frac{1}{\Gamma^{(n)}} \mbox{Tr}_\mathbb{B} \left[UP^{(n)} \gamma_0 P^{(n)}U^\dagger\right] .
\end{equation}

Is this process given by a linear map?  Dynamically, the  evolution of the total state $\gamma_0$ is  linear.  However, for the purpose of tomography, the dynamics of the prepared input states $P^{(n)}$ is of interest,  but $P^{(n)}$ appears twice in the process equation, therefore the output states $Q^{(n)}$ depend bi-linearly on $P^{(n)}$. That is suppose $P=\alpha A+\beta B$, where $A$ and $B$ are some operators and $\alpha$ and $\beta$ are constants then
\begin{eqnarray*}
P\gamma_0 P&=&(\alpha A+\beta B)\gamma_0(\alpha A+\beta B)\\
&=&\alpha^2 A\gamma_0A+\alpha \beta A\gamma_0B+\alpha \beta B\gamma_0A+\beta^2 B\gamma_0B.
\end{eqnarray*}
The last result shows that the $Q$ has a bi-linear dependence on $P$. This bi-linearity can also be seen from the dependence of the environment state on the state of the system.  To see this expanding Eq. (\ref{BasicProcessEquation}) as:
\begin{eqnarray*}
Q^{(n)} &=&  \mbox{Tr}_\mathbb{B} \left[U P^{(n)} \otimes \tau^{(n)} U^\dagger\right],\\
\mbox{with}\hspace{.5cm}&\\
\tau^{(n)} &=& \frac{1}{\Gamma^{(n)}} \mbox{Tr}_\mathbb{A}\left[P^{(n)}\gamma_0\right].
\end{eqnarray*}
The last equation clearly shows that the environment state $\tau^{(n)}$ is in effect a function of $P^{(n)}$.  It is well known \cite{CarteretTernoZyczkowski05, RomeroEtal04, Ziman06} that if the initial state of the system is related to the state of the environment by some function $f$:
\begin{equation*}\label{nonlinev}
\rho^\mathbb{AB} = \sum_j \rho^\mathbb{A}_j \otimes f(\rho^\mathbb{A}_j)^{\mathbb B},
\end{equation*}
where where $\rho^{\mathbb A}$ is the state of the system and $f(\rho^{\mathbb{A}})^{\mathbb B}$ are the density matrix in the space of the environment.  Then the evolution of the reduced matrices $\rho^\mathbb{A}$ cannot be consistently described by a single linear map.   In this case, the function $f$ is of a specific form that gives us a bi-linear dependence.

We now look at a simple example to demonstrate that when input states are prepared by measurement, the results cannot be consistently described by a linear map. 

%%%%%%%%%%%%%%%%%%%%%%%%%%%%%%%%%%%%%%%%
\subsection{Example of Preparation by Measurements}\label{mesex}
%%%%%%%%%%%%%%%%%%%%%%%%%%%%%%%%%%%%%%%%

Our example demonstrates when the output states are bi-linearly related to the input states the standard quantum process tomography procedure will give incorrect results.  Suppose we are unaware of the bi-linear dependence that arises from preparing input states by measurements, and assume that the process is given by a linear map.  We would prepare a set of linearly independent input states, then construct the linear process map from the measured output states and the duals of input states.  We will show that this linear map will not give the correct prediction for certain inputs. 

Consider the following two qubit state as the available state to the experimenter at $t=0_-$:
\begin{eqnarray}\label{totst}
\gamma_0=\frac{1}{4}\left(\openone\otimes\openone+a_j\sigma_j\otimes\openone+c_{23}\sigma_2\otimes\sigma_3\right).
\end{eqnarray}
We will treat the first qubit as the system and the second qubit as the environment.  Note that the the total state is separable, but it is correlated with the environment.

Once again we will use the input states given in Eq. (\ref{prj}). The state of the system plus the environment after each measurement takes the following form:
\begin{eqnarray}\label{mesinputs1}
P^{(n)}\gamma_0P^{(n)}\rightarrow P^{(n)}\otimes\frac{1}{2}\openone\;\;(\mbox{for}\;n=\{(1,\pm),(3,+)\} ),\nonumber\\
P^{(2,+)}\gamma_0P^{(2,+)}\rightarrow P^{(2,+)}\otimes\frac{1}{2}\left(\openone+\frac{c_{23}}{1+a_2}\sigma_3\right).\hspace{.5cm}
\end{eqnarray}

Following the same recipe as before the output states are obtained using Eq. (\ref{BasicProcessEquation}). We will use the same unitary $U$ as in the last example given in Eq. (\ref{unitary}).  The output states are as follow:
\begin{eqnarray*}
Q^{(1,-)}&=&\frac{1}{2}\left(\openone-C^2\sigma_1\right),\;\;
Q^{(1,+)}=\frac{1}{2}\left(\openone+C^2\sigma_1\right),\\
Q^{(2,+)}&=&\frac{1}{2}\left(\openone-c^{+}_{23}CS\sigma_1+C^2\sigma_2+c^{+}_{23}S^2\sigma_3\right),\\
\mbox{and} && Q^{(3,+)}=\frac{1}{2}\left(\openone+ S^2\sigma_3\right),
\end{eqnarray*}
where $C=\cos(2t)$, $S=\sin(2t)$, and $c^{+}_{23}=\frac{c_{23}}{1+a_2}$.  

The linear process map is constructed using Eq. (\ref{qptlin}), the duals in Eq. (\ref{dual}), and the output states:
\begin{eqnarray*}
\small{\Lambda_{m}=\frac{1}{2}\left(\begin{array}{cccc}
1+C^2 & ic^{+}_{23} S^2 &0 & 2C^2-ic^{+}_{23}C S  \vspace{.2 cm}\\
-ic^{+}_{23}S^2 & 1-C^2 & ic^{+}_{23}C S & 0 \vspace{.2 cm}\\
0 & -ic^{+}_{23}C S &1-C^2 & -ic^{+}_{23}S^2  \vspace{.2 cm}\\
 2C^2+ic^{+}_{23}C S & 0& ic^{+}_{23}S^2 & 1+C^2 \vspace{.2 cm}
\end{array}\right).}
\end{eqnarray*}

Now consider the action of $\Lambda_m$ on state $P^{(2,-)}=\frac{1}{2}(\openone-\sigma_2)$.  $P^{(2,-)}$ is a linear combination of the input states as $P^{(2,-)}=P^{(1,-)}+P^{(1,+)}-P^{(2,+)}$.  If the action of $\Lambda_m$ is linear then the output state corresponding to the input state $P^{(2,-)}$, should be 
\begin{eqnarray}\label{lact}
\Lambda_m\left(P^{(2,-)}\right)&=&\Lambda_m\left(P^{(1,-)}+P^{(1,+)}-P^{(2,+)}\right)\nonumber\\
&=&Q^{(1,-)}+Q^{(1,+)}-Q^{(2,+)}\\
&=&\frac{1}{2}\left(\openone+c^{+}_{23}CS\sigma_1
-C^2\sigma_2-c^{+}_{23}S^2\sigma_3\right).\nonumber
\end{eqnarray}

Is this the same as if the global state was prepared in the state $P^{(2,-)}$ by a measurement?  The output state for input $P^{(2,-)}$ can be calculated using Eq. (\ref{BasicProcessEquation}),
\begin{eqnarray}\label{realst}
Q^{(2,-)}&=&\frac{1}{\Gamma_{(2,-)}}\mbox{Tr}_{\mathbb B}\left[UP^{(2,-)}\gamma_0P^{(2,-)}U^{\dag}\right]\nonumber\\
&=&\frac{1}{2}\left(\openone-c^{-}_{23} CS\sigma_1-C^2\sigma_2- c^{-}_{23}S^2\sigma_3\right),
\end{eqnarray}
where $c^{-}_{23}=\frac{c_{23}}{1-a_2}$.  

Clearly the output state predicted by the linear process map in Eq. (\ref{lact}) is not the same as the real output state calculated dynamically in Eq. (\ref{realst}), hence the linear process map $\Lambda_m$ does not describe the process correctly.  This is not surprising, the state of the environment in Eq. (\ref{mesinputs1}) depends on  $a_2$, and subsequently  the linear process map depends on $a_2$, hence on the state of the system.  This is where the non-linearity of the process arises.

In next section we will show that this process can be correctly described
using a bi-linear process map.  We will also develop a new tomography
procedure for bi-linear process maps.

%%%%%%%%%%%%%%%%%%%%%%%%%%%%%%%%%%%%%%%%
%%%%%%%%%%%%%%%%%%%%%%%%%%%%%%%%%%%%%%%%
\section{Characterization of Bi-linear Quantum Processes}\label{bilin}
%%%%%%%%%%%%%%%%%%%%%%%%%%%%%%%%%%%%%%%%
%%%%%%%%%%%%%%%%%%%%%%%%%%%%%%%%%%%%%%%%

Let us expand Eq. (\ref{BasicProcessEquation}) with matrix indices:
\begin{eqnarray}\label{RawProcessEquation}
Q^{(n)}_{r,s} 
& =& \frac{1}{\Gamma^{(n)}} U_{r \epsilon,r' \alpha} P^{(n)}_{r'r''} {\gamma_0}_{r''\alpha,s''\beta} P^{(n)}_{s''s'} U^*_{s \epsilon,s' \beta} \nonumber\\
& =& \frac{1}{\Gamma^{(n)}} {\tensor*{{P}}{*^{(n)^*}_{r''r'}}}
\left(U_{r \epsilon,r' \alpha} {\gamma_0}_{r''\alpha,s''\beta}  U^*_{s \epsilon,s' \beta} \right)P^{(n)}_{s''s'} \nonumber\\
& =& \frac{1}{\Gamma^{(n)}} {\tensor*{{P}}{*^{(n)^*}_{r''r'}}}M^{(r,s)}_{r''r';s''s'} P^{(n)}_{s''s'}\;.
\end{eqnarray}
In the last equation, the matrix $M$ is defined as:
\begin{equation}\label{mmat}
M^{(r,s)}_{r''r'; s''s'} = \sum_{\alpha \beta \epsilon} U_{r \epsilon,r' \alpha} 
{\gamma_0}_{r''\alpha,s''\beta} U^*_{s \epsilon,s' \beta} .
\end{equation}
Note that in Eq. \ref{RawProcessEquation} the superscript indices on $M$ match the elements on the left hand side of the equation, while the subscript indices are summed on the right hand of the equation. 

The output state $Q^{(n)}$ is given by the matrix $M$ acting bi-linearly on the input state $P^{(n)}$.  Therefore the matrix $M$ fully describes the process, and we will call $M$ \emph{the bi-linear process map}.

$M$ contains both $U$ and $\gamma_0$, however knowing $M$, is not sufficient to determine $U$ and $\gamma_0$.  As expected, it should not be possible to determine $U$ and $\gamma_0$ through measurements and preparations on the system alone, without access to the environment.  Conversely, $M$ contains all the information that is necessary to fully determine the output state for any prepared input state.

Before proceeding to determine the matrix $M$, let us look at some basic properties of $M$.

%%%%%%%%%%%%%%%%%%%%%%%%%%%%%%%%%%%%%%%%
\subsection{Some Basic Properties of $M$}
%%%%%%%%%%%%%%%%%%%%%%%%%%%%%%%%%%%%%%%%

Let us start with the trace of $M$.  To take the trace equate indices $r$ with $s$, $r'$ with $s'$, and $r''$ with $s''$ to get:
\begin{eqnarray*}
\mbox{Tr}[M]&=&\sum_{rr'r''} M_{r''r',r''r'}^{(r,r)}\\
& =& \sum_{ \alpha \beta \epsilon}\sum_{rr'r''}
U_{r \epsilon,r' \alpha} {\gamma_0}_{r''\alpha,r''\beta}U^*_{r \epsilon,r' \beta}.
\end{eqnarray*}
Since $U^\dag U=I\Rightarrow U^*_{r\epsilon,r'\beta}U_{r\epsilon,r'\alpha}=\delta_{\alpha\beta}$, then
\begin{eqnarray*}
\mbox{Tr}[M]& =&\sum_{\alpha r''} {\gamma_0}_{r''\alpha,r''\alpha}
=1 .
\end{eqnarray*}

As with the case of linear process maps, matrix $M$ is hermitian.  This is easy to see by taking the complex conjugate of $M$,
\begin{eqnarray*}
\left(M_{r''r',s''s'}^{(r,s)}\right)^* &=&\left( \sum_{\alpha \beta \epsilon} U_{r \epsilon,r' \alpha} 
{\gamma_0}_{r''\alpha,s''\beta} U^*_{s \epsilon,s' \beta} \right)^*\\
&=&\sum_{\alpha \beta \epsilon}  U_{s \epsilon,s' \beta}
{\gamma_0}_{s''\beta ,r''\alpha}U^*_{r \epsilon,r' \alpha,} \\
&=&M_{s''s',r''r'}^{(s,r)}.
\end{eqnarray*}
The complex conjugate of $M$ is not only the transpose of $M$ but each element of $M$ is also transposed.

%%%%%%%%%%%%%%%%%%%%%%%%%%%%%%%%%%%%%%%
\subsection{Bi-linear map for a qubit}\label{twotomo}
%%%%%%%%%%%%%%%%%%%%%%%%%%%%%%%%%%%%%%%

We now have to develop a new tomography procedure to deal with the bi-linear process map $M$.  We will need to figure out a finite set of input states  and the corresponding output states that will allow us to determine $M$.  Afterall, this is the objective of tomography -- by performing measurements on a small number of select input states a complete description of a process can be obtained and predict the output state for \emph{any} arbitrary input state.

$M$ is a large ($N^3 \times N^3)$ matrix, where $N$ is the dimension of the space of system. To make it more manageable, Eq. (\ref{RawProcessEquation}) can be interpreted as:
\begin{equation}
\Brak{P^{(n)}}{M}{P^{(n)}}= \Gamma^{(n)} Q^{(n)},
\label{SimplifiedProcessEquation}
\end{equation}
where $P^{(n)}$ is now treated as a vector.  In this form, $M$ is a $N^2 \times N^2$ matrix with each element being a $N \times N$ matrix.  We will call Eq (\ref{SimplifiedProcessEquation}) the `\emph{bi-linear process equation}'.  

Since $M$ is a Hermitian matrix, it has $\frac{1}{2}(N^4+N^2)$ independent block elements.  Therefore $\frac{1}{2}(N^4+N^2)$ independent equations in the form of Eq. (\ref{SimplifiedProcessEquation}) are necessary to fully determine $M$.  It is clear that neither an orthonormal set of $N$ or linearly independent set of $N^2$ input states would provide sufficient number of equations to resolve the bi-linear process map $M$.

Consider a qubit system.
In that case the projections $P^{(n)}$ can be written in terms its coherence vector $a_j$ and Pauli spin matrices $\sigma_j$:
\begin{equation}
P^{(n)} = \frac{1}{2} \left(\openone +\sum_{j=1}^3 a^{(n)}_j \sigma_j\right).
\label{PVectorDecomposition}
\end{equation}
Since $P^{(n)}$ is a projection then there is an additional constrain $\sum_j ({a^{(n)}_j})^2=1$.

The matrices $\openone$ and $\sigma_j$ together forms a vector basis for this space.  Therefore Eq. (\ref{PVectorDecomposition}) is simply a vector decomposition of $P^{(n)}$ in a fixed basis.
Taking this form for $P^{(n)}$ and substituting into Eq. (\ref{SimplifiedProcessEquation}) gives:
\begin{eqnarray}
\Gamma^{(n)} Q^{(n)}&=&\bra{P^{(n)}}M\ket{P^{(n)}}\nonumber\\
&=&\frac{1}{4}\braket{\openone | M | \openone } +\frac{a^{(n)}_j}{4} \left(\braket{ \openone | M | \sigma_j } +\braket{ \sigma_j | M | \openone } \right) \nonumber\\
&&+\frac{a^{(n)}_j a^{(n)}_k}{4} \braket{ \sigma_j | M | \sigma_k }.
\label{OpenQubitProcessEquation}
\end{eqnarray}

Observe that the terms $\braket{ \openone | M | \openone }$, $\braket{ \openone | M | \sigma_j }$, $\braket{ \sigma_j | M | \openone }$ and $\braket{ \sigma_j | M | \sigma_k }$, are simply the matrix elements of $M$ in $\{ \openone, \sigma_j\}$ basis.  We just need to find a set of projections $P^{(n)}$ that will allow us to solve for these matrix elements.

Consider the following specific projections defined as $P^{(j,\pm)}=\frac{1}{2} (\openone \pm \sigma_j)$ with $j=\{1,2,3\}$.
\begin{eqnarray}
\Gamma^{(j,\pm)}Q^{(j,\pm)}&=& \braket{P^{(j,\pm)}|M|P^{(j,\pm)}}\nonumber\\
&=&\frac{1}{4}(\braket{ \openone | M | \openone }+ \braket{ \sigma_j | M | \sigma_j } )\\ 
&& \hspace{.4cm}\pm\frac{1}{4}( \braket{ \sigma_j | M | \openone } + \braket{ \openone | M | \sigma_j }) .\nonumber
\end{eqnarray}
Simultaneously solving the $(+)$ and $(-)$ equations above give the following unknowns:
\begin{eqnarray*}
\braket{ \openone | M | \openone}+ \braket{ \sigma_j | M | \sigma_j  }=2\left(\Gamma^{(j,+)}Q^{(j,+)}+\Gamma^{(j,-)}Q^{(j,-)}\right)\\
\braket{ \openone | M | \sigma_j } + \braket{ \sigma_j | M | \openone }=2\left(\Gamma^{(j,+)}Q^{(j,+)}-\Gamma^{(j,-)}Q^{(j,-)}\right).
\end{eqnarray*}

To obtain the cross terms $\braket{ \sigma_j | M | \sigma_k }$ consider projections such as $P^{\left(j+k+1,+\right)}=\frac{1}{2} \left(\openone + \frac{1}{\sqrt{2}} \sigma_j + \frac{1}{\sqrt{2}} \sigma_k\right)$ for $k>j$ which gives:
\begin{eqnarray}
\Gamma^{(j+k+1,+)}Q^{(j+k+1,+)}&=& \bra{P^{(j+k+1,+)}}M\ket{P^{(j+k+1,+)}}\nonumber\\
&=&\frac{1}{8}\left(\braket{ \openone | M | \openone } + \braket{ \sigma_j | M | \sigma_j } \right)\nonumber \\
&&+\frac{1}{8}\left(\braket{ \openone | M | \openone }+  \braket{ \sigma_k | M | \sigma_k }\right)\nonumber \\
&&+ \frac{1}{4\sqrt{2}} \left(\braket{ \openone | M | \sigma_j } + \braket{ \sigma_j | M | \openone }\right) \nonumber\\
&&+ \frac{1}{4\sqrt{2}} \left(\braket{ \openone | M | \sigma_k } + \braket{ \sigma_k | M | \openone } \right)\nonumber\\
&& +\frac{1}{8}\left(\braket{ \sigma_j | M | \sigma_k } +  \braket{ \sigma_j | M | \sigma_k }\right).\nonumber
\end{eqnarray}
Substitute the known terms and solve for the desired cross terms,
\begin{eqnarray}
\braket{ \sigma_j | M | \sigma_k }+  \braket{ \sigma_k | M | \sigma_j }
=-2\left(1+ {\sqrt{2}}\right)\Gamma^{(j,+)}Q^{(j,+)} \nonumber\\
-2\left(1-{\sqrt{2}}\right)\Gamma^{(j,-)}Q^{(j,-)}\nonumber \\
-2\left(1+\sqrt{2}\right)\Gamma^{(k,+)}Q^{(k,+)} \nonumber\\
-2\left(1-\sqrt{2}\right)\Gamma^{(k,-)}Q^{(k,-)} \nonumber\\
+8\Gamma^{(j+k+1,+)}Q^{(j+k+1,+)}\nonumber.
\end{eqnarray}

In summary using the following nine projections,
\begin{eqnarray}\label{choice1}
P^{(j,+)}=\frac{1}{2} \left(\openone + \sigma_j\right),\;\;
P^{(j,-)}=\frac{1}{2} \left(\openone - \sigma_j\right) ,\nonumber \\
P^{(4,+)}=\frac{1}{2} \left(\openone + \frac{1}{\sqrt{2}} \sigma_1 + \frac{1}{\sqrt{2}} \sigma_2\right) ,\nonumber \\
P^{(5,+)}=\frac{1}{2} \left(\openone + \frac{1}{\sqrt{2}} \sigma_1 + \frac{1}{\sqrt{2}} \sigma_3\right) ,\\
P^{(6,+)}=\frac{1}{2} \left(\openone + \frac{1}{\sqrt{2}} \sigma_2 + \frac{1}{\sqrt{2}} \sigma_3\right),\nonumber
\end{eqnarray}
and solving them simultaneously yields all desired matrix elements:
$\braket{ \openone | M | \openone } + \braket{ \sigma_j | M | \sigma_j }$,
 $\braket{ \openone | M | \sigma_j} + \braket{ \sigma_j | M | \openone }$,
 and $\braket{ \sigma_j | M | \sigma_k } + \braket{ \sigma_k | M | \sigma_j}.$

Though this is not enough to fully determine $M$, these elements are sufficient to determine the output state for any input state. Using the property $\sum_j (a^{(n)}_j)^2= 1$, Eq. (\ref{OpenQubitProcessEquation}) can be rewritten as:
\begin{eqnarray}\label{M_Elements}
4 \Gamma^{(n)} Q^{(n)} &=&
\sum_j \left(a^{(n)}_j\right)^2 \left(\braket{ \openone | M | \openone } + \braket{ \sigma_j | M | \sigma_j } \right) \nonumber\\
&&+ \sum_j a^{(n)}_j \left( \braket{ \openone | M | \sigma_j } + \braket{ \sigma_j | M | \openone } \right) \\
&&+ \sum_{k>j} a^{(n)}_j a^{(n)}_k \left( \braket{ \sigma_j | M | \sigma_k } + \braket{ \sigma_k | M | \sigma_j } \right).\nonumber
\end{eqnarray}

Observe that the sums of the cross terms $\braket{ \sigma_j | M | \sigma_k } + \braket{ \sigma_k | M | \sigma_j }$ can appear together because the coefficients $a^{(n)}_j$ are real.  Also, the element $\braket{ \openone | M | \openone }$ can always be paired with a diagonal element $\braket{ \sigma_j | M | \sigma_j }$ as long as the state is a pure projection satisfying $\sum_j (a^{(n)}_j)^2= 1$.  The diagonal element $\braket{ \openone | M | \openone }$ only has to be known if the system can be prepared directly to a mixed state such that $\sum_j (a^{(n)}_j)^2< 1$.  This may be accomplished by a generalized measurement \cite{Kuah02} (see appendix \ref{genmes}).  If generalized  measurements are allowed, then just one more input state is needed, for example $\frac{1}{2}( \openone + \frac{1}{2} \sigma_1)$, which gives another independent equation that can be solved to obtain $\braket{ \openone | M | \openone }$. 

Therefore the elements of $M$ found in Eq. (\ref{M_Elements}) are all that are needed to describe the process. By measuring the outputs for the nine specified input states, the matrix $M$ can be calculated. We now have a good quantum process tomography procedure for an open two-level system.

This procedure can be generalized for an $n$-level system in the same spirit as above.  However care must be taken, since the constraints on the coherence vector of an $n$-level system are more complicated (see \cite{byrd:062322} for a detailed discussion).
 
%Note that the nine states used above is not a unique choice.  The recipe which used to derive these nine states can be used in principle to derive other choices, and can also be partly generalized to $N$-level systems.  However, there are some non-trivial obstacles to overcome for the generalization to $N$-level systems.  In place of the Pauli matrices for two-level systems, the generalized Pauli-Gellman hermitian traceless $N \times N$ matrices can be used \cite{Greiner} to decompose the $N \times N$ density matrix, and this decomposition will also have only real coefficients.  This trick eliminates certain degrees of freedom in the matrix $M$ that is otherwise difficult to deal with.  Unfortunately, for $N > 2$, these real coefficients no longer satisfy just the simple constraint $\sum_j (a^{(n)}_j)^2= 1$.  The additional constraints on the coefficients complicate the task of constructing the projections needed to simultaneously span the matrix elements of $M$.

%%%%%%%%%%%%%%%%%%%%%%%%%%%%%%%%%%%%%%%%
%%%%%%%%%%%%%%%%%%%%%%%%%%%%%%%%%%%%%%%%
\section{Bi-linear vs. linear process map verification procedure}\label{bvl}
%%%%%%%%%%%%%%%%%%%%%%%%%%%%%%%%%%%%%%%%
%%%%%%%%%%%%%%%%%%%%%%%%%%%%%%%%%%%%%%%%

In practice it is not easy to tell if a process is linear or bi-linear. The bi-linear process map is incompatible with the behavior of a linear process map and conversely a linear process map will not adequately describe a bi-linear process. We have developed a procedure below that will differentiate between a linear process and a bi-linear process.

Consider what happens to a state that is a linear combination of a set of projections:
\begin{equation}\label{unknowstate}
X = \sum_n c_n P^{(n)} .  
\end{equation}
If the evolution of $X$ though the process is bi-linear then the output is written in terms of the bi-linear process map as:
\begin{equation}
\braket{ X | M | X } =\sum_{mn} c_m c_n \bra{ P^{(m)}}M \ket{P^{(n)} }.
\label{BilinearMapUnknown}
\end{equation}
But if the bi-linear process map is compatible in some way with a linear process, then Eq. (\ref{BilinearMapUnknown}) can be simplified to
\begin{equation}\label{unknowstate3}
\braket{ X | M | X } = \sum_n c_n^2 \bra{ P^{(n)}} M  \ket{P^{(n)} } .
\end{equation}
In that case the two equations above are equal:
\begin{equation}
\sum_{mn} c_m c_n \bra{ P^{(m)}}M \ket{P^{(n)} } = \sum_n c_n^2 \braket{P^{(n)} | M | P^{(n)} } .
\label{BilinearMapContradiction}
\end{equation}
It is clear that no non-trivial conditions exists for $M$ that will allow this equality for arbitrary coefficients $c_m$. Therefore the bi-linear process map gives different predictions from a linear process map.  In that case, we should be able to distinguish between whether a process is given by a linear map or a bi-linear map.  

Consider the example of the tomography procedure proposed for a two-level system.  If the process is given by a linear process map, then the nine input states in Eq. (\ref{choice1}) are over complete; only four input states are needed to determine a linear process map.  This discrepancy is summarized by the following linear sum rules:
\begin{eqnarray*}\label{linrules}
P^{(2,-)}&=&  P^{(1,+)} + P^{(1,-)}-P^{(2,+)},\nonumber\\
P^{(3,-)}&=&  P^{(1,+)} + P^{(1,-)}-P^{(3,+)},\\
P^{(4,\pm)} &=& \frac{1}{2}\left(P^{(1,+)} + P^{(1,-)}\right)
\pm \frac{1}{\sqrt{2}} \left(P^{(2,+)}-P^{(1,-)}\right),\\
P^{(5,\pm)} &=& \frac{1}{2}\left(P^{(1,+)} + P^{(1,-)}\right)
\pm \frac{1}{\sqrt{2}} \left(P^{(3,+)}-P^{(1,-)}\right),\\
P^{(6,\pm)} &=& \frac{1}{2}\left(P^{(1,+)} + P^{(1,-)}\right)\\
&&\pm \frac{1}{\sqrt{2}} \left(P^{(2,+)}+P^{(3,+)}-P^{(1,+)}-P^{(1,-)}\right).
\end{eqnarray*}

If the process is linear, then the output states must satisfy the same sum rules, which are obtained from the above equations by suitably writing $Q$ in place of $P$.  If these sum rules are not satisfied, then the process is not linear.  

However satisfying the sum rules is necessary but not sufficient to determine if the process is linear; a bi-linear process map can still be constructed from this set of input and output states without contradiction.  Therefore an additional input state, distinct from the above nine input states, should be tested to determine which of Eq. (\ref{BilinearMapUnknown}) and Eq. (\ref{unknowstate3}) is satisfied.  

More explicitly, if the output state corresponding to an additional input state of the form of Eq. (\ref{unknowstate}) is found to be $\sum_n c_n^2 \Gamma^{(n)}Q^{(n)}$, then the process is linear.  However if the corresponding output state is given by $\sum_{mn} c_m c_n \braket{P^{(m)} | M | P^{(n)} }$ then the process is bi-linear.

%%%%%%%%%%%%%%%%%%%%%%%%%%%%%%%%%%%%%%%
%%%%%%%%%%%%%%%%%%%%%%%%%%%%%%%%%%%%%%%
\section{Fundamental issues of state preparation beyond quantum process tomography}\label{fun}
%%%%%%%%%%%%%%%%%%%%%%%%%%%%%%%%%%%%%%%
%%%%%%%%%%%%%%%%%%%%%%%%%%%%%%%%%%%%%%%

Although our discussion has largely focused on quantum process tomography, we would like to emphasize that the process equations describe \emph{any} quantum experiment.  Before any experiment begins, the quantum system or particle would exist in an unknown state that could be (and most likely is) correlated with a quantum environment.  Preparation of the system or particle into a known state is a necessary part of any experiment.  

As we have shown, when the system or particle is imperfectly isolated, in other words, it interacts non-trivially with a quantum environment during the experiment, the initial step of state preparation in an experiment is of fundamental importance.  

We described two methods of state preparation, the stochastic preparation and the preparation by measurements.  For preparation by measurements, the outcomes of the experiment will be non-linearly related to the prepared states.  So it would seem that the stochastic method is preferable, since with the stochastic method, the evolution of the \emph{prepared states} to the final states is linear.  However, how the stochastic maps are actually performed may need to be carefully considered.  

Stochastic maps can be equivalently performed by a unitary transformation with the addition of ancillary quantum systems.  In particular, the way the stochastic preparation is constructed in section \ref{stoprep} is equivalent to a generalized measurement (see appendix \ref{genmes}).  

Recall that with the stochastic preparation, different maps $\mathscr{P}^{(n)} = \Omega^{(n)} \circ \Theta$  prepare different input states.  Let $K$ be the number of input states.  Assume that for the experiment, an equal number of each input state is prepared, so the expectation map is:
\begin{equation*}
\sum_{n=1}^K \frac{1}{K} \mathscr{P}^{(n)}.
\end{equation*}

This overall, expectation map is a trace preserving map since each $\mathscr{P}^{(n)}$ preserves trace.  Therefore this can be considered to be a generalized measurement, where each $\frac{1}{K} \mathscr{P}^{(n)}$ represents a measurement outcome.

A generalized measurement can be equivalently performed by a unitary transformation and a von Neumann measurement by using a suitable ancillary system.  How does this implementation of the stochastic preparation effect our results?  Let the ancillary system be labeled as $\mathbb{C}$, and the generalized measurement is implemented with a unitary transformation $W$ and von Neumann measurement given by the projections $J^{(n)}$; where the $J^{(n)}$ outcome corresponds with the $n^{th}$ preparation map $\frac{1}{K} \mathscr{P}^{(n)}$:
\begin{equation}
\mathscr{P}^{(n)} (\rho^\mathbb{A}) = \mbox{Tr}_\mathbb{C} \left[{J^{(n)}}^\mathbb{C} W^\mathbb{AC} \rho^\mathbb{A} \otimes \epsilon^\mathbb{C} {W^\mathbb{AC}}^\dagger {J^{(n)}}^\mathbb{C}\right].
\end{equation}
where $\epsilon^\mathbb{C}$ is the initial state of the ancillary system.

Now include the above equation within the overall process equation Eq. (\ref{stocproceq})
\begin{equation*}
{Q^{(n)}}
= \mbox{Tr}_\mathbb{BC} \left[ U {J^{(n)}}^\mathbb{C} W^\mathbb{AC} \gamma_0^\mathbb{AB} \otimes \epsilon^\mathbb{C} {W^\mathbb{AC}}^\dagger {J^{(n)}}^\mathbb{C} U^\dagger \right] .
\end{equation*}

If $U$ acts only on the system $\mathbb{A}$ and environment $\mathbb{B}$, and not on the ancillary $\mathbb{C}$, then the above equation can be simplified:
\begin{eqnarray}
{Q^{(n)}}
& = &\mbox{Tr}_\mathbb{B}\left[U \mbox{Tr}_\mathbb{C} \left[{J^{(n)}}^\mathbb{C} W^\mathbb{AC} \gamma_0^\mathbb{AB} \otimes \epsilon^\mathbb{C}  {W^\mathbb{AC}}^\dagger {J^{(n)}}^\mathbb{C}\right] U^\dagger \right] \nonumber\\
& = &\mbox{Tr}_\mathbb{B} \left[ U {\mathscr{P} ^{(n)}}^\mathbb{A} \otimes \mathcal{I}^\mathbb{B} \left(\gamma_0^\mathbb{AB}\right) U^\dagger \right] \\
& =& \mbox{Tr}_\mathbb{B} \left[U {P^{(n)}}^\mathbb{A} \otimes \tau^\mathbb{B} U^\dagger\right] . \nonumber
\end{eqnarray}
And the result is the linear process map as before.  However, if $U$ acts non-trivially on our system $\mathbb{A}$, environment $\mathbb{B}$, and the ancillary $\mathbb{C}$ then we cannot make this simplification, and $Q^{(n)}$ would have a bi-linear dependence on $J^{(n)}$, and in turn a bi-linear dependence on the prepared input states.  

We can make two important observations from this.  First, the quantum environment should be defined as any quantum system that interacts with the primary system $\mathbb{A}$ \emph{during} the experiment.  The quantum environment \emph{does not} have to include any quantum systems that the primary system is entangled or correlated with, if there is no interaction between the two during the experiment.

The second observation is that if the stochastic preparation method is used, any ancillary systems used to implement the stochastic maps, must be perfectly isolated from the primary system during the experiment.  In other words, they must not be part of the quantum environment.  If the ancillary systems used are not properly isolated, then the process may have bi-linear dependence on the input states.

This is an important result -- simply deciding on a stochastic preparation method for an experiment may not in practice guarantee that the process will be linear.  This can have implications in many areas, in particular, quantum error correction protocols are designed to correct for linear noise, correcting for bi-linear effects may be a more difficult challenge.

The verification procedure discussed in the last section may therefore be important, since it can be used as a tool to confirm the proper isolation of the apparatus and ancillary systems during the experiment.  We can verify in practice if a process is linear instead of simply making that assumption.  In the following section, we will describe a complete experiment that may be performed with qubits, including verification steps to check if the process is linear or bi-linear.

%%%%%%%%%%%%%%%%%%%%%%%%%%%%%%%%%%%%%%%%
%%%%%%%%%%%%%%%%%%%%%%%%%%%%%%%%%%%%%%%%
\section{A complete recipe for an experiment}\label{rec}
%%%%%%%%%%%%%%%%%%%%%%%%%%%%%%%%%%%%%%%%
%%%%%%%%%%%%%%%%%%%%%%%%%%%%%%%%%%%%%%%%

Although we have established the theory, let us make the ideas more concrete by developing a complete recipe for an experiment that can be used to determine whether a process is bi-linear or linear.  We will also show specifically how the corresponding bi-linear map or linear map can be calculated from the measurement results.

For bi-linear process tomography, nine input states are necessary.  For the nine states derived in section \ref{twotomo} the first six states are three pairs of orthonormal projections, but the last three are not.  Now consider twelve projections, nine from Eq. (\ref{choice1}) and three orthogonal to the last three in that equation:
\begin{eqnarray}
P^{(4,-)} = \frac{1}{2} \left(\openone - \frac{1}{\sqrt{2}} \sigma_1 - \frac{1}{\sqrt{2}} \sigma_2\right) , \nonumber\\
P^{(5,-)} = \frac{1}{2} \left(\openone - \frac{1}{\sqrt{2}} \sigma_1 - \frac{1}{\sqrt{2}} \sigma_3\right) , \\
P^{(6,-)} = \frac{1}{2} \left(\openone - \frac{1}{\sqrt{2}} \sigma_2 - \frac{1}{\sqrt{2}} \sigma_3\right) \nonumber.
\end{eqnarray}

These twelve projections are neatly grouped into six different sets of orthonormal pairs.  If the states are prepared using von Neumann measurements, these would correspond to measurements in the $\sigma_1$, $\sigma_2$, $\sigma_3$, $\sigma_1 + \sigma_2$, $\sigma_1 + \sigma_3$ and $\sigma_2 + \sigma_3$ directions.  Twelve states are more than necessary for bi-linear process tomography, but the extra states can utilize as consistency checks.

After recording the corresponding output states for all twelve input states, the linearity of the process can be verified. If the process is linear then only four linearly independent input states are necessary to determine it.  The other eight input states can be written in linear combination of these four.  If the states from Eq. (\ref{prj}) are used as the four linearly independent states then the following eight linear sum rules (that the input states satisfy) have to be satisfied:
\begin{eqnarray*}
Q^{(2,-)}&=&  Q^{(1,+)} + Q^{(1,-)}-Q^{(2,+)},\\
Q^{(3,-)}&=&  Q^{(1,+)} + Q^{(1,-)}-Q^{(3,+)},\\
Q^{(4,\pm)} &=& \frac{1}{2}\left(Q^{(1,+)} + Q^{(1,-)}\right)
\pm \frac{1}{\sqrt{2}} \left(Q^{(2,+)}-Q^{(1,-)}\right),\\
Q^{(5,\pm)} &=& \frac{1}{2}\left(Q^{(1,+)} + Q^{(1,-)}\right)
\pm \frac{1}{\sqrt{2}} \left(Q^{(3,+)}-Q^{(1,-)}\right),\\
Q^{(6,\pm)} &=& \frac{1}{2}\left(Q^{(1,+)} + Q^{(1,-)}\right)\\
&&\pm \frac{1}{\sqrt{2}} \left(Q^{(2,+)}+Q^{(3,+)}-Q^{(1,+)}-Q^{(1,-)}\right).
\end{eqnarray*}
If the eight sum rules are satisfied, then the process is not bi-linear, and should be described by a linear process map.  The linear process map can then be computed following the recipe in section \ref{LQPT}.
%\begin{eqnarray}
%\Lambda\left(\frac{1}{2}(\openone + p_j \sigma_j)\right) &=&Q^{(1,+)}+ Q^{(1,-)}\nonumber\\
%&&+ p_j \left(Q^{(j,+)} - Q^{(j,-)}\right).
%\end{eqnarray}

If the eight sum rules are not satisfied, then the process may be given by a bi-linear map.  We will attempt to verify that the process is bi-linear and calculate the bi-linear process map.

Note that the probabilities $\Gamma^{(n)} = \mbox{Tr}[\gamma_0 P^{(n)}]$ associated with each preparation should be found experimentally.  The probabilities should be complete for an orthonormal set of projections, in other words:
$\Gamma^{(j,+)} + \Gamma^{(j,-)} = 1.$
Therefore the probabilities can be calculated from the fraction of the $(+)$ states as compared to the $(-)$ states, for all preparations made in the same direction.

If the process is bi-linear then three additional states should evolve in a way that is consistent with a bi-linear map derived from the other nine states.  The following equations are derived from this condition and using Eq. (\ref{M_Elements}).  If these equations are satisfied then the process is bi-linear:
\begin{eqnarray*}
\Gamma^{(4,-)} Q^{(4,-)}
&=&\frac{1}{\sqrt{2}}\left(
\Gamma^{(1,-)} Q^{(1,-)}-\Gamma^{(1,+)} Q^{(1,+)}\right.\\
&&\;\;\;\; \left.+\Gamma^{(2,-)} Q^{(2,-)}-\Gamma^{(2,+)} Q^{(2,+)}\right)\\
&&\;\;\;\;\;\;+\Gamma^{(4,+)} Q^{(4,+)},\\
\Gamma^{(5,-)} Q^{(5,-)}
&=&\frac{1}{\sqrt{2}}\left( \Gamma^{(1,-)} Q^{(1,-)}-\Gamma^{(1,+)} Q^{(1,+)}\right.\\
&&\;\;\;\; \left. + \Gamma^{(3,-)} Q^{(3,-)}- \Gamma^{(3,+)} Q^{(3,+)}\right)\\
&&\;\;\;\;\;\;+\Gamma^{(5,+)} Q^{(5,+)},\\
\Gamma^{(6,-)} Q^{(6,-)}
&=&\frac{1}{\sqrt{2}}\left( \Gamma^{(2,-)} Q^{(2,-)}-\Gamma^{(2,+)} Q^{(2,+)}\right.\\
&&\;\;\;\; \left. + \Gamma^{(3,-)} Q^{(3,-)}- \Gamma^{(3,+)} Q^{(3,+)}\right)\\
&&\;\;\;\;\;\;+\Gamma^{(6,+)} Q^{(6,+)}.
\end{eqnarray*}
If the conditions above are satisfied then the process is bi-linear, and the bi-linear process map can be computed by following the recipe in section \ref{twotomo}.

Once the matrix elements of $M$ are determined in this fashion, the evolution of any state $X = \frac{1}{2}(\openone + \sum_j p_j \sigma_j)$ is given by:
\begin{eqnarray*}
 4\Gamma Q&=& 4\braket{X | M | X }\\
&=& \sum_j p_j^2 \left(\braket{\openone | M | \openone } + \braket{\sigma_j | M | \sigma_j } \right)\\
&&+\sum_j p_j \left( \braket{\openone | M | \sigma_i } + \braket{\sigma_j | M | \openone } \right)\\
&&+\sum_{k>j} p_j p_k \left( \braket{\sigma_j | M | \sigma_k } + \braket{\sigma_k | M | \sigma_j } \right) .
\end{eqnarray*}

Note that since $Q$ is a normalized state, the normalization constant $\Gamma$ is the measurement probability $\Gamma = \mbox{Tr}[X \gamma_0]$.  Although we had not explicitly mentioned this before, the matrix $M$ contains all information about the measurement probabilities, that is why we needed the measurement probabilities $\Gamma^{(n)}$ to calculate the matrix $M$.

Finally note that if both the test for linearity and bi-linearity fails, then the process cannot be consistently described by either a linear map or bi-linear map.  The experiment then should be carefully analyzed for problems such as any non-linear dependence that may have been introduced if the input states are not accurately prepared, or if there is some dependence of $\gamma_0$ on the prepared state.

%%%%%%%%%%%%%%%%%%%%%%%%%%%%%%%%%%%%%%%
%%%%%%%%%%%%%%%%%%%%%%%%%%%%%%%%%%%%%%%
\section{Analysis of a quantum process tomography experiment}\label{exper}
%%%%%%%%%%%%%%%%%%%%%%%%%%%%%%%%%%%%%%%
%%%%%%%%%%%%%%%%%%%%%%%%%%%%%%%%%%%%%%%

In this section we will analyze a quantum process tomography experiment performed by M. Howard et al. \cite{Howard06}.  Our critique will emphasize the importance of having a consistent theory of state preparation.

In this experiment, the system that is studied is an electron configuration formed in a nitrogen vacancy defect in a diamond lattice.  The quantum state of the system is given by a spin triplet (S=1).  Again we will write the initial state of the system and environment as $\gamma_0$.  

The system is prepared by optical pumping, which results in a strong spin polarization.  The state of the system is said to have 70\% chance of being in a pure state $\ket{\phi}$.  Or more mathematically, the probability of $\ket{\phi}$ is $\mbox{Tr}[\ket{\phi}\bra{\phi} \gamma_0] = 0.7$.  

Since the population probability is high, an assumption was made that the state of the system can be simply approximated as a pure state $\ket{\phi}\bra{\phi}$.  From this initial state, different input states can be prepared by suitably applying microwave pulses resonant with the transition levels.  After preparation, the system is allowed to evolve, and the final states (density matrices) are measured.  With the knowledge of the initial state and the measured final states, the linear process map that should describe this process is determined.  

It was found that the liner process map has negative eigenvalues, so the map was ``corrected" using a least squares fit between the experimentally determined map and a theoretical map based on Hermitian parametrization \cite{havel03}, while enforcing complete positivity. 

However, if we do not regard the negative eigenvalues of the map as aberrations, then we should consider the assumptions about the preparation of the system more carefully.  The assumption about the initial state of the system is:
\begin{equation}
\gamma_0 \rightarrow \ket{\phi}\bra{\phi}\otimes\tau .
\end{equation}
This is in effect a pin map.  Together with the stochastic transformation of the initial state into the various input states, this is identical to the stochastic preparation method discussed in section \ref{stoprep}.

It is clear that the pure initial state assumption is unreasonable, given our knowledge now of how the process is sensitive to the initial correlations between the system and the environment.  In effect the action of the pin map in this experiment is not perfect, and therefore the pin map can be ignored.  Then the process equation is:
\begin{equation}
Q^{(n)} = \mbox{Tr}_\mathbb{B} [U \Omega^{(n)} \otimes \openone (\gamma_0) U^\dagger] 
\end{equation}
where $\Omega^{(n)}$ is the stochastic mapping corresponding to preparing the $n^{th}$input state.  We will assume that the stochastic process does not involve any ancillary systems that interact with the primary system during the experiment.

In this experiment, $\Omega^{(n)}$ is nothing more than a unitary transformation $V^{(n)}$ satisfying $V^{(n)} \ket{\phi} = \ket{\psi^{(n)}}$, where $\ket{\psi^{(n)}}$ is the desired pure $n^{th}$ input state to the process.

We can write the unitary transformation for a two-level system as:
\begin{equation}
V^{(n)} = \ket{\psi^{(n)}}\bra{\phi}+ \ket{\psi^{(n)}_\perp }\bra{\phi_\perp}
\end{equation}
where $\braket{\psi^{(n)}|\psi^{(n)}_\perp } = 0$ and $\braket{\phi | \phi_\perp } = 0$.  This basically defines $V^{(n)}$ as a transformation from the basis $\{\ket{\phi}\}$ to the basis $\{\ket{\psi^{(n)}_i}\}$.  The equation for the process becomes:
\begin{eqnarray*}
Q^{(n)} &=& \mbox{Tr}_\mathbb{B} \left[U \ket{\psi^{(n)}}\braket{\phi| \gamma_0 |\phi}\bra{\psi^{(n)}} U^\dagger\right]\\
&& +  \mbox{Tr}_\mathbb{B} \left[U \ket{\psi^{(n)}_\perp}\braket{\phi_\perp| \gamma_0 |\phi}\bra{\psi^{(n)}} U^\dagger\right]\\
&& +  \mbox{Tr}_\mathbb{B} \left[U \ket{\psi^{(n)}}\braket{\phi| \gamma_0 |\phi_\perp}\bra{\psi^{(n)}_\perp} U^\dagger\right]\\
&& +  \mbox{Tr}_\mathbb{B} \left[U \ket{\psi^{(n)}_\perp}\braket{\phi_\perp| \gamma_0 |\phi_\perp}\bra{\psi^{(n)}_\perp} U^\dagger\right].
\end{eqnarray*}
Therefore, since $\braket{\phi | \gamma_0 | \phi} = 0.7$, to first approximation the process is a linear mapping on the states $\ket{\psi^{(n)}}\bra{\psi^{(n)}}$.  However it is clear that if all terms are included, the process is not truly a linear map of the states $\ket{\psi^{(n)}}\bra{\psi^{(n)}}$.  The negative eigenvalues are therefore a result of fitting results into a linear process map when the process is not truly  linear.

%%%%%%%%%%%%%%%%%%%%%%%%%%%%%%%%%%%%%%%
%%%%%%%%%%%%%%%%%%%%%%%%%%%%%%%%%%%%%%%
\section{Conclusions}\label{con}
%%%%%%%%%%%%%%%%%%%%%%%%%%%%%%%%%%%%%%%
%%%%%%%%%%%%%%%%%%%%%%%%%%%%%%%%%%%%%%%

Preparation of the system or particle into a known state is a necessary part of any experiment.  We have shown that with open systems, some care has to be taken to define the method of state preparation.

We described two methods, the stochastic method and the measurement method.  The stochastic method leads linear processes that can be described by linear process maps.  The advantage of the measurement method is that the primary system does not have to be perfectly isolated. This method only requires a good measurement apparatus.

With the stochastic method, the initial state can be made simply separable, effectively de-coupling the system from the environment.  The evolution of the system is then given by a linear process map.  However we find that the isolation of the apparatus from the system during the experiment is of greater importance with this method.  Any apparatus or ancillary systems used for the stochastic preparation must not be contained in the quantum environment, the quantum environment being defined as everything that interacts with the quantum system during the experiment.

The stochastic method is more consistent with the traditional method of performing quantum experiments, and its advantage is that the process map is effectively equivalent to the linear dynamical map.  However the disadvantage to the stochastic method is that the apparatus and any ancillary systems employed to perform the stochastic transformations or generalized measurements, must be perfectly isolated from the primary system for the duration of the experiment.

With the measurement method, if the initial state is correlated with the environment then the evolution of such a system is given by a bi-linear process map.  The determination of this bi-linear process map by process tomography is more difficult, but we developed a procedure that works for qubit systems.

If interaction occurs between the system and this environment, then non-linear noise can be introduced into the experiment.  This could have fundamental consequences, for example quantum error correction schemes proposed so far are based on correcting linear noise.  If bi-linear errors occur, error correction becomes a more difficult challenge.

We proposed a protocol to distinguish between a bi-linear process and linear process.  This protocol can be used to verify the assumptions made about state preparation in the experiment, by adding some additional inputs to the experiment and making consistency checks, the correctness of a stochastic preparation can be \emph{verified}.  The protocol we proposed can be an practical experimental tool.

\begin{acknowledgments}
We would like to thank Anil Shaji and Thomas Jordan for useful discussions that lead us to consider the problem of quantum process tomography for open systems. We thank Ken Shih and Pablo Bianucci for many discussions about quantum process tomography experiments.
\end{acknowledgments}

\appendix

%%%%%%%%%%%%%%%%%%%%%%%%%%%%%%%%%%%%%%%
%%%%%%%%%%%%%%%%%%%%%%%%%%%%%%%%%%%%%%%
\section{generalized measurements}\label{genmes}
%%%%%%%%%%%%%%%%%%%%%%%%%%%%%%%%%%%%%%%
%%%%%%%%%%%%%%%%%%%%%%%%%%%%%%%%%%%%%%%

%%%%%%%%%%%%%%%%%%%%%%%%%%%%%%%%%%%%%%%
%%%%%%%%%%%%%%%%%%%%%%%%%%%%%%%%%%%%%%%
\subsection{Overview of measurements}
%%%%%%%%%%%%%%%%%%%%%%%%%%%%%%%%%%%%%%%
%%%%%%%%%%%%%%%%%%%%%%%%%%%%%%%%%%%%%%%

Measurements in quantum theory begin with the von Neumann measurement \cite{vonNeumann32}, which can be quickly summarized as follows:  if the system being measured is in the state $\rho$ and the measurement is given by a set of orthonormal projections $\Pi_j$, then the probability of obtaining the $j^{th}$ outcome is $\mbox{Tr}[\rho \Pi_j]$.  If the $j^{th}$ outcome is observed, then the state collapses to $\Pi_j$.

An early generalization of the von Neumann measurement was given \cite{Jauch67,Davies70}, with the introduction of POVMs.  With POVMs, the measurement is still given by a set of operators $\Xi_j$, however the operators need not be orthonormal projections, but are in general positive operators satisfying $\sum_j \Xi_j = 1$.  The probability of obtaining the $j^{th}$ outcome is still $\mbox{Tr}[\rho \Xi_j]$ and the state collapses to $\frac{\Xi_j}{\mbox{Tr}[\Xi_j]}$.

However, measurements can be generalized beyond POVMs (see \cite{Kuah02} for more discussions).  Suppose a very elaborate measurement apparatus is constructed, one that involves many transformations and POVMs.  Rather than to look at all the intricate details of this apparatus the apparatus can be treated as a black box.  When a state is fed into the box one of several possible outcomes is registered, and a new state $\rho'$ leaves the box, which is fixed for that particular outcome.

The motivation is to use the most general linear operation on a density matrix as a measurement, a linear map.  A general measurement can be framed as follows: given a quantum system in state $\rho$, let the measurement be given by a set of positive trace reducing linear maps $\mathcal{B}^{(j)}$.  The probability of registering the $j^{th}$ outcome is given by $\mbox{Tr}[\mathcal{B}^{(j)} (\rho)]$.  If the $j^{th}$ outcome is observed, then the system collapses to the state:
\begin{equation}
\rho \rightarrow \rho'=\frac{\mathcal{B}^{(j)} (\rho)}{\mbox{Tr}[\mathcal{B}^{(j)} (\rho)]}.
\end{equation}
To ensure the probabilities sum to one, the sum of the maps $\sum_i \mathcal{B}^{(j)}$ must be itself a trace preserving map.  

Writing $\mathcal{B}^{(j)}$ its canonical form with its eigen-matrices $C^{(j)}_\alpha$ and eigenvalues $c^{(j)}_\alpha$ yields:
\begin{equation}
\mathcal{B}^{(j)} (\rho) = \sum_\alpha c^{(j)}_\alpha 
C^{(j)}_\alpha \rho \;{C^{(j)}_\alpha}^\dagger.
\end{equation}
The condition that the overall map $\sum_j \mathcal{B}^{(j)}$ preserves trace gives: 
\begin{equation}
\sum_{j \alpha} c^{(j)}_\alpha {C^{(j)}_\alpha}^\dagger
 {C^{(j)}_\alpha}= \openone.
\label{measurementmapssumcondition}
\end{equation}

Given this condition, we can prove that this very general scheme can be accomplished with a unitary transformation and von Neumann measurement, by using an ancillary system of sufficient size.  The method we use here is similar to the method used to show how a trace preserving map can be given as a reduced unitary transformation \cite{Sudarshan86}.

%%%%%%%%%%%%%%%%%%%%%%%%%%%%%%%%%%%%%%%
%%%%%%%%%%%%%%%%%%%%%%%%%%%%%%%%%%%%%%%
\subsection{Generalized measurements}\label{constgenmes}
%%%%%%%%%%%%%%%%%%%%%%%%%%%%%%%%%%%%%%%
%%%%%%%%%%%%%%%%%%%%%%%%%%%%%%%%%%%%%%%

Consider a transformation $W$ that acts on a tensor product space consisting of the original system and two ancillary systems.  Let the original system and the ancillary systems be spanned by the basis states $\ket{r,j,\alpha}$ respectively.  The action of $W$ is defined as follows:
\begin{equation}
W: \ket{r',0,0} \rightarrow \sum_{rj\alpha} 
\sqrt{c^{(j)}_\alpha} [C^{(j)}_\alpha]_{rr'} 
\ket{r,j,\alpha}.
\label{Vdef}
\end{equation}
The size of the ancillary systems is bounded by $\mu N^2$, since $j$ ranges from $1$ to $\mu$, where $\mu$ is the number of maps $\mathcal{B}^{(j)}$ making up the measurement and $\alpha$ ranges from $1$ to $N^2$ (where $N$ is dimension of the system) since each map $ \mathcal{B}^{(j)}$ has at most $N^2$ $C$-matrices.

The action of the transformation $W$ on a complete set of basis states is not yet defined. However for the states on which it is defined, $W$ does preserve orthornormality between those states.  The proof is straight forward using Eq. \ref{measurementmapssumcondition}:
\begin{eqnarray*}
\left(\bra{r',0,0} W^\dagger \right) 
\left(W\ket{s',0,0} \right)
&=& \sum_{r j \alpha} c^{(j)}_\alpha
[C^{(j)}_\alpha]^*_{r'r} [C^{(j)}_\alpha]_{rs'}\nonumber \\
&=& \delta_{r's'}.
\end{eqnarray*}
Since $W$ preserves the orthonormality, it can be made into a valid unitary transformation by carefully defining its action on the remaining space that is so far not covered by equation \ref{Vdef}. 

Now we demonstrate that our generalized measurement, given by the set of maps $\mathcal{B}^{(j)}$, can be equivalently performed by this unitary transformation $W$ and a von Neumann measurement.  Performing the unitary transformation $W$ on the original system in state $\rho$ and the ancillary systems in the initial state $\ket{0,0}\bra{0,0}$ to gives:
\begin{eqnarray}
\chi &=& W \left(\rho \otimes \ket{0,0}\bra{0,0} \right) W^\dagger \nonumber\\
%&=& W \left(\rho_{r's'} \ket{r',0,0}\bra{s',0,0} \right) W^\dagger \nonumber\\
& =& \sum_{rsr's'jk \alpha \beta}
\sqrt{c^{(j)}_\alpha c^{(k)}_\beta}
[C^{(j)}_\alpha]_{rr'}
\rho_{r's'}
[C^{(k)}_\beta]^*_{s's}\\
&&\hspace{1.5cm}
\times\ket{r, j, \alpha}\bra{ s,k,\beta}. \nonumber
\end{eqnarray}

Now perform a von Neumann measurement on the first ancillary system, given by the set of orthonormal projections $\ket{j}\bra{ j}$. The probability of the $j^{th}$ outcome is:
\begin{eqnarray}
\mbox{Tr}[\chi \ket{j}\bra{ j} ]
&=& \sum_{rr's'\alpha} c^{(j)}_\alpha
[C^{(j)}_\alpha]_{rr'}\rho_{r' s'}[C^{(j)}_\alpha]^*_{s'r}\\ 
&=& \mbox{Tr}[\mathcal{B}^{(j)} (\rho)]. \nonumber
\end{eqnarray}

If the $j^{th}$ outcome is observed, then the original plus ancillary system collapses to the state:
\begin{eqnarray*}
\bra{ j}\chi \ket{j}&=&\frac{1}{K} \sum_{r s r's'   \alpha \beta}
\sqrt{c^{(j)}_\alpha
c^{(j)}_\beta}
[C^{(j)}_\alpha]_{rr'}
\rho_{r' s'}
[C^{(j)}_\beta]^*_{s's}\\
&&\hspace{1.5cm}
\times\ket{r, j, \alpha}\bra{ s,j,\beta}.
\end{eqnarray*}
where $K = \mbox{Tr}[\mathcal{B}^{(n)} (\rho)]$ normalizes the state.

Finally the measurement is over and the particle exits the apparatus.  The state of the system, now outside the apparatus, is given by tracing over the remaining ancillary system inside the apparatus:
\begin{eqnarray*}
\rho_{r's'} \rightarrow \rho'_{rs}&=&\frac{1}{K} \sum_{rsr's'\alpha}
c^{(n)}_\alpha [C^{(n)}_\alpha]_{rr'}\rho_{r' s'}
[C^{(n)}_\alpha]^*_{ss'}\\
&=& \frac{[\mathcal{B}^{(n)} (\rho)]_{rs}}{ \mbox{Tr}[\mathcal{B}^{(n)} (\rho)]}.
\end{eqnarray*}

Therefore this gives the same results as the generalized measurement we laid out.  We only needed an ancillary system big enough (dimension $\mu N^2$), one unitary transformation, and one von Neumann measurement to perform the most general quantum measurement.

%%%%%%%%%%%%%%%%%%%%%%%%%%%%%%%%%%%%%%%%
%%%%%%%%%%%%%%%%%%%%%%%%%%%%%%%%%%%%%%%%
\section{Process vs. Dynamical Maps}\label{pvd}
%%%%%%%%%%%%%%%%%%%%%%%%%%%%%%%%%%%%%%%%
%%%%%%%%%%%%%%%%%%%%%%%%%%%%%%%%%%%%%%%%

Dynamical maps \cite {SudarshanMatthewsRau61,SudarshanJordan61} allow the description of stochastic processes and the evolution of open systems.  Dynamical maps used to describe the evolution of open systems are usually defined with a constant environment state $\tau$:
\begin{eqnarray*}
\mathscr{B}\left(\rho^{\mathbb A}\right) = \mbox{Tr}_\mathbb{B}\left[U \rho \otimes \tau U^\dagger\right]
\end{eqnarray*}
Implicitly, the state of the environment $\tau$ is a parameter of the map $\mathscr{B}$.  Therefore, the linear dynamical map $\mathscr{B}$ would only consistently describe an experiment if different input states $\rho$ can be prepared independently of the environment state $\tau$.  The actual issue of how this can be executed is never addressed.

The stochastic preparation method provides a way for an experiment to be made so that different input states can be prepared with a fixed environment state.  Consider the process equation Eq. (\ref{stocproceq}) in section \ref{stoprep}.
\begin{eqnarray}
Q^{(n)} &=& \mbox{Tr}_\mathbb{B} \bigg{[} U \left.\left[\Omega^{(n)} \circ\Theta\right] \otimes \mathcal{I} \right. \left(\gamma_0\right) U^\dagger\bigg{]}\nonumber \\
& =& \mbox{Tr}_\mathbb{B}\left[U P^{(n)} \otimes \tau(\Theta) U^\dagger\right].\nonumber
\end{eqnarray}
The process map is then given by:
\begin{eqnarray}
\Lambda\left(\rho^{\mathbb A} \right) &=& \mbox{Tr}_\mathbb{B} \bigg{[} U \rho^{\mathbb A} \otimes \tau(\Theta) U^\dagger\bigg{]}\nonumber
\end{eqnarray}
Therefore, in this context, the dynamical map is equivalent to the process map.  However, for consistency, we have to remember that the environment state is a constant to the problem, therefore the pin map $\Theta$ should also be a constant to the problem.

It is possible to consider a dynamical map where the environment is not fixed, such as the reduced dynamical evolution of a non-simply separable state $\gamma_0$ \cite{JordanShajiSudarshan04}:
\begin{eqnarray}\label{dmaps}
\mathscr{B}\left(\mbox{Tr}_{\mathbb B}\left[\gamma_0\right]\right)=\mbox{Tr}_{\mathbb B}\left[U\gamma_0U^\dag\right] 
\end{eqnarray}
The dynamical map in this problem is applicable only over a compatibility domain of states, rather than over the complete state space of the system $\mathbb{A}$.  The compatibility domain \cite{JordanShajiSudarshan04,PhysRevA.64.062106} is the set of states that are compatible with the correlations in $\gamma_0$.  Formally, this problem defines an extension map\cite{PhysRevA.64.062106,breuer:022103,unp} (also known as preparations) that relates the initial state of the system $\mathbb{A}$ to the overall initial state of $\mathbb{AB}$.  The extension map is linear but not necessarily a completely positive map.  Therefore, in this context, such dynamical maps are a theoretical tool, and have a limited relation to a physical experiment.  This why we differentiate between process maps and dynamical maps.

\bibliography{tomop.bib}

\end{document}